# Photofragmentation Dynamics and Dissociation Energies of MoO and CrO

Graham A. Cooper, Alexander S. Gentleman, Andreas Iskra, and Stuart R. Mackenzie*

Department of Chemistry, University of Oxford, Physical and Theoretical Chemistry Laboratory, South Parks Road, Oxford, OX1 3QZ, United Kingdom

ABSTRACT

Neutral metal-containing molecules and clusters present a particular challenge to velocity map imaging techniques. Common methods of choice for producing such species – such as laser ablation or magnetron sputtering – typically generate a wide variety of metal-containing species and, without the possibility of mass-selection, even determining the identity of the dissociating moiety can be challenging. In recent years we have developed a velocity map imaging spectrometer equipped with laser ablation source explicitly for studying neutral metal-containing species. Here, we report the results of velocity map imaging photofragmentation studies of MoO and CrO. In both cases, dissociation at the two– and three– photon level leads to fragmentation into a range of product channels, some of which can be confidently assigned to particular Mo* (Cr*) and O atom quantum states. Analysis of the kinetic energy release spectra as a function of photon energy allows precise determination of the ground state dissociation energies of MoO ( = $44064 \pm 133$ cm$^{-1}$) and CrO (= $37197 \pm 78$ cm$^{-1}$), respectively.

# I. INTRODUCTION

Neutral transition metal oxides (TMOs) have been the subject of intense spectroscopic and thermochemical interest as a result of their fundamental importance in areas as diverse as catalysis, biology and astrophysics.[1-4] Various important geometric and electronic properties of neutral diatomic TMOs have been obtained, including equilibrium bond lengths, rotational constants, vibrational frequencies, and electronic term energies.[1, 2, 5-7] Despite this wealth of information, some key thermochemical values such as ground state dissociation energies ($D_0$) remain poorly determined. Traditionally, dissociation energies for metal oxides (MOs) are obtained from MO ionization energies in combination with MO$^+$ bond energies (from mass spectrometric studies) and well-known metal atom ionization potentials *via* the simple thermochemical cycle:

$$D_0(\text{MO}^+) + IE(\text{MO}) = D_0(\text{MO}) + IE(\text{M}). \tag{1}$$

As a result, the precision of the $D_0$(MO) values obtained is limited to that of the least well-determined value. Although increasingly amenable to modern photoelectron spectroscopy methods,[8, 9] the majority of literature values for the MO ionization energies come from electron impact ionization studies which suffer large experimental uncertainties.

Ion imaging,[10] and particularly the velocity map imaging (VMI) technique,[11] permit the photodissociation dynamics of neutral molecules to be investigated with unprecedented resolution and can yield $D_0$ values with much improved uncertainties compared with conventional spectroscopic methods.

In recent years in Oxford, we have developed a velocity map imaging spectrometer equipped with a laser ablation source for the study of metal–containing species. The application of VMI to neutral (metal–containing) molecules is, however, non–trivial. Laser ablation typically generates a wide range of molecules, radicals and clusters. As a result, even with selective photofragment ionization / detection, the inability to mass–select the parent species means it can be challenging to identify the molecule from which a particular fragment originates. In some cases it has been possible to take advantage of clear spectroscopic signatures to identify the molecular carrier, e.g., Au-RG,[12, 13] $Cu_2$,[14] CuO,[14] $Au_2$,[15] $Xe_2$,[16, 17] $Li(NH_3)_4$.[18] In other examples (such as Ag–RG, AgO and $Ag_2$[19]), the identity of the co–fragment has been obtained from the dependence of the kinetic release upon photon energy. Particularly relevant to the current paper, these VMI studies, with their direct measurement of dissociation, have led to improved values for the experimental dissociation energies of AgO and CuO, over previous literature values.[14, 19] Here, we report an extension of these studies to the photodissociation dynamics of MoO and CrO.

The dissociation energy of MoO has been previously studied by a variety of techniques. Early studies relied on the Knudsen cell mass spectrometry, from which values of 40600 ± 5200 cm$^{-1}$ by DeMaria *et al.*,[20] and 46550 ± 1800 cm$^{-1}$ by Emelyanov *et al.*[21] were determined. More recently, the thermochemical cycle approach of Equation 1 has been used following high–resolution photoelectron spectroscopy determinations of the MoO ionization energy. Loock *et al.*,[8] and Luo *et al.*,[9] determined values for $D_0$(MoO) of 43900 ± 300 cm$^{-1}$, and 43700 ± 200 cm$^{-1}$, respectively. The difference between these values arises from the different dissociation energies of MoO$^+$ used. Luo *et al.* relied on guided ion beam studies by Sievers *et al.*,[22] while Loock *et al.* used a combination of this value and another determined by Kretzschmar *et al.* by a combined mass spectrometry and computational study.[23]

In contrast to MoO, a much wider range of CrO studies have resulted in poor agreement on the ground state dissociation energy. Knudsen cell measurements by Balducci *et al.* produced a value of 36510 ± 750 cm$^{-1}$,[24] while two guided ion beam studies using a similar thermochemical cycle to that mentioned above for MoO, reported values of 39000 ± 700 cm$^{-1}$ and 36900 ± 600 cm$^{-1}$.[25, 26] A 1993 thermodynamic review by Ebbinghaus gives a value of $D_0$(CrO) of 38800 cm$^{-1}$,[27] while more recent measurements of 38200 ± 700 cm$^{-1}$ (using crossed molecular beam reactivity[28]) and 38510 cm$^{-1}$ (from reaction kinetics[29]) have been reported.

Here, we report the results of multiphoton dissociation dynamics of MoO and CrO in the visible and the near UV studied by velocity map imaging. Extrapolation of the observed kinetic energy release as a function of photoexcitation energy yields improved ground state dissociation energies which are compared with existing literature values.

## II. EXPERIMENTAL

The instrument and basic methodology used for these VMI experiments has been described previously.[12, 14, 15] In brief, pulsed 532 nm laser ablation of a rotating disc target of natural molybdenum or chromium metal (natural isotopic abundance) is used to generate Mo or Cr–containing species which are entrained in a pulsed molecular beam of pure argon or helium. For these studies, two photoexcitation sources have been used: Visible excitation in the 445 – 475 nm region (22470 – 21050 cm$^{-1}$) was provided by the output of an optical parametric oscillator (Continuum Panther, 8ns pulses), and tuneable ultraviolet laser pulses from a frequency–doubled dye laser (Sirah). In both cases, the same pulse of light both induces fragmentation and ionizes excited electronic states of metal atom dissociation fragments. In all experiments, photoexcitation takes places within the extraction region of a VMI spectrometer, arranged with ion extraction collinear with the molecular beam.[14]

Images were recorded by detecting $Mo^+$ or $Cr^+$ ions produced by photoionization of Mo* or Cr* atomic photofragments. Optimal image resolution is achieved by gating over with the isotope with the highest natural abundance ($^{52}Cr^+$ or $^{98}Mo^+$). Images were recorded at a 10 Hz repetition rate for typically 30–60 minutes in order to achieve sufficient signal intensity for meaningful analysis. Reconstruction of the central slice of the ion sphere was performed using either the Maximum Entropy Velocity Image Reconstruction (MEVIR) method,[30] or the Polar Onion Peeling (POP) algorithm.[31]

Following ablation of the Mo target, the main species observed in the molecular beam upon photoexcitation were $Mo^+$ and $MoO^+$, with traces of $MoO_2^+$ and $Mo_2^+$. The presence of neutral MoO in the beam was confirmed by recording the 1 + 2 REMPI spectrum in the 21200 to 22200 $cm^{-1}$ region for comparison with that previously reported by Hamrick *et al.*.[32] These resonances, however, do not appear to be involved in the dissociation processes studied here as no enhancement in the photofragment yield was observed when the probe laser was resonant with the bands recorded.

Ablation of the Cr target, coupled with 455 nm multiphoton ionization, yields significant $Cr^+$ and $CrO^+$, with $Cr_2^+$ signals suggesting the presence in the beam of the corresponding neutrals. 2 + 1 REMPI spectra of Cr atoms were observed between 21200 and 23600 $cm^{-1}$, but no assignable CrO spectrum was recorded.

**III. RESULTS AND DISCUSSION**

*A. Multiphoton dissociation of MoO in the Visible Region*

Figure 1 shows some representative velocity map images recorded in the region around 21700 cm$^{-1}$, together with a selection of Mo* fragment Kinetic Energy Release (fKER) spectra recorded in this spectral region. These images are obtained by gating over the $^{98}$Mo$^+$ ion signal. Although this corresponds to only *ca*. 24% of the atomic ions produced (all isotopes being detected in proportion to their natural abundance), this yields optimal VMI resolution.

The images show three main rings which are present in each case, as well as a number of weaker, broader features present in some of the images. In most VMI photodissociation studies, the parent molecule is known due to the use of stable precursors[33] or (for charged species) mass-selectivity.[34, 35] This is not the case here due to our use of laser ablation for producing a range of neutral Mo–containing species in the molecular beam. The identity of the co-fragment (and hence that of the signal carrier) can, however, be determined from the evolution of the VMI rings with excitation wavenumber. For two-body photodissociation of the MoX molecule at the *n* photon level (*i.e.*, MoX + *n*hν → Mo* + X) conservation of energy dictates that

$$E_{\text{MoX}} + nh\nu = D_0(\text{Mo-X}) + E_{\text{Mo}} + E_{\text{X}} + \text{TKER}, \qquad (2)$$

where $E_{\text{MoX}}$, $E_{\text{Mo}}$, and $E_{\text{X}}$ are the internal energies of the parent molecule, the molybdenum atom photofragment and the co-fragment, respectively, $D_0(\text{Mo–X})$ is the dissociation energy and TKER is the total kinetic energy release. Since, in the centre of mass frame, the momenta of the Mo and X fragments must be equal and opposite, the TKER can be inferred from the kinetic energy of the Mo fragment alone, fKER, and by the ratio of masses, such that

$$\text{fKER} = \frac{p_{\text{Mo}}^2}{2m_{\text{Mo}}} = \frac{m_X}{m_{\text{MoX}}} \cdot \text{TKER} \propto nh\nu \,. \tag{3}$$

Thus, by determining how the fKER increases with photoexcitation wavenumber we can determine both the mass of the co-fragment, $m_x$, and the number of photons, $n$, absorbed before dissociation. Here, this approach yields two plausible combinations of $m_X$ and $n$ and thus for the dissociation process involved. The first of these involves single-photon dissociation of $Mo_2$. However, this can be ruled out on energetic grounds as the $Mo_2$ bond energy is known to be around 36000 cm$^{-1}$,[36, 37] much higher than the photon energy used. The alternative process, photodissociation of MoO at the three photon level is the preferred assignment and the basis for the following analysis.

Figure 2 shows the variation in the total kinetic energy release as a function of photon wavenumber for all dissociation channels observed. Three distinct series (numbered Series 1, 2 and 3 in Figure 2) are observed. The dataset used includes several images in which all three channels are observed. These provide independent measurements of the TKER separation between the three channels yielding average separations of 1596 ± 218 cm$^{-1}$ (between Series 3 and 2), and 1893 ± 247 cm$^{-1}$ (Series 1 and 2). These compare well to the 1609 cm$^{-1}$ and 1801 cm$^{-1}$ energy gaps between central spin-orbit components of the $^5G$ (at 16747 cm$^{-1}$), $^5P$ (~18356 cm$^{-1}$) and $^5D$ (~20157 cm$^{-1}$) terms arising from the $4d^55s$ configuration of Mo.[38] On this basis, the observed dissociation channels can be confidently assigned as:

Series 1: $^{98}$MoO ($^5\Pi$) + 3 $h\nu$ → Mo* ($4d^55s$, $b\ ^5D$) + O ($^3P$)

Series 2: $^{98}$MoO ($^5\Pi$) + 3 $h\nu$ → Mo* ($4d^55s$, $a\ ^5P$) + O ($^3P$)

Series 3: $^{98}$MoO ($^5\Pi$) + 3 $h\nu$ → Mo* ($4d^55s$, $a\ ^5G$) + O ($^3P$) (4)

The spin-orbit components of these levels (which in the case of the $b\,^5D$ state span 381 cm$^{-1}$) and of the oxygen $^3P_J$ ground state are not resolved in this study and thus account for some of the width of the peaks in the TKER spectra. Uncertainty in the internal energies of the fragments resulting from this splitting has been included in the error analysis of these results. In each series, following dissociation at the three-photon level, a further two photons are required to ionise the Mo* fragment before detection is possible.

Having assigned the dissociation thresholds, an extrapolation of the best-fit lines through each of the three series back to zero TKER yields a precise measure of the dissociation threshold relative to the MoO ground state. Each series yields an independent measure of $D_0(^{98}\text{MoO})$ from which a weighted average of 44064 ± 133 cm$^{-1}$ is obtained (see Table 1).

### B. Ultraviolet photodissociation of MoO

Figure 3 shows typical velocity map images recorded following photoexcitation in the ultraviolet region around 312 nm (32050 cm$^{-1}$), along with their corresponding fragment kinetic energy release spectra. In these images all Mo$^+$ isotopes were detected simultaneously to increase the count rate.

Three distinct dissociation channels are again observed, the rings for which demonstrate wavelength-dependence as indicated. The same procedure as that in Section IIIA was used to determine the nature of the parent molecule. Only two plausible assignments emerge: single photon dissociation of MoAr, or dissociation of MoO at the two photon level. In this case, the former cannot be discounted on energetic grounds. However, assuming a likely dissociation energy of < 200 cm$^{-1}$ for MoAr, single-photon photolysis at these energies would leave > 30000 cm$^{-1}$ in Mo* electronic excitation. The few dissociation channels observed in the VMI images (at most three clear rings are observed) is inconsistent with the high density of Mo* terms (and thus open dissociation channels) at this energy. Fitting assuming MoO dissociation is more convincing.

Figure 4 shows the TKER spectra as a function of photon wavenumber assuming MoO fragmentation. Again three main series (labelled I, II, and III) can be identified and, on the basis of the match between the energetic separation of these series and those between low-lying Mo* states, the series are assigned as:

Series I: $^{98}$MoO ($^5\Pi$) + 2 $h\nu$ → Mo* ($4d^5 5s$, $a\ ^5P$) + O ($^3P$)

Series II: $^{98}$MoO ($^5\Pi$) + 2 $h\nu$ → Mo* ($4d^5 5s$, $a\ ^5G$) + O ($^3P$)

Series III: $^{98}$MoO ($^5\Pi$) + 2 $h\nu$ → Mo* ($4d^4 5s^2$, $a\ ^5D$) + O ($^3P$)   (5)

The different spin-orbit components of the fragment terms produced are not resolved again leading to broad TKER features. This is most significant in the case of Series III, where the 1381 cm$^{-1}$ difference in wavenumber between the $J = 0$ and 4 levels of the $a\ ^5D$ term results in a particularly broad TKER peak. By way of comparison, the narrowest individual TKER peaks observed in this study have FWHM $ca.$ 500 cm$^{-1}$. These images therefore represent two–photon dissociation into two of the same channels involved in the three–photon dissociations in Section IIIA. Figure 5 illustrates these two cases schematically, with the position of the levels involved indicated. In each case the excited states accessed lie just above the MoO ionization energy. However, predissociation to neutral fragments appears to compete effectively with the ionization process. Fitting assuming photodissociation of the $MoO^+$ cation was attempted but the results were less satisfactory.

Once again, values for $D_0(MoO)$ can be determined from each series by extrapolation to TKER = 0, and these are given, along with their weighted average of $43895 \pm 148$ cm$^{-1}$, in Table 1. This value lies within 0.5% of that determined in Section IIIA thus providing supporting evidence for the three-photon results. The $44064 \pm 133$ cm$^{-1}$ value is considered the more reliable given that it was derived from experiments employing isotope-selective detection.

The MoO dissociation energy determined here is consistent with both the older Knudsen cell measurements,[5, 6, 20, 21] and the newer values reported using the ionisation energies of MoO.[8, 9] In contrast to previous values, however, our $D_0(MoO)$ value is directly observed and is not dependent on other species. This avoids the problems suffered by Loock *et al.*[8] and Luo *et al.*[9] who determined values for the ionization energy within 3 cm$^{-1}$ of each other, but differ in their final values for $D_0(MoO)$ by 200 cm$^{-1}$ as a result of assuming different values for the dissociation energy of $MoO^+$. Our value is slightly smaller than that calculated by Oliveira *et al.* who reported a value of *ca*. 45000 cm$^{-1}$ for $D_e$ though approximately half of the difference can be accounted for by the zero-point energy, *ca*. 450 cm$^{-1}$.[32]

*C. Dissociation of CrO in the Visible Region*

CrO represents the first-row $d^5$ analogue of MoO and, to facilitate comparison of the two, images were recorded at similar photon energies to those in Section IIIA, following ablation of a chromium target. Figure 6 shows sample images and fKER spectra gated on $^{52}$Cr$^+$ isotope. The kinetic energy release spectra are more complex than observed for MoO but two clear dissociation thresholds are observed in all of the images, along with several other, less persistent features.

The VMI data can be understood in terms of three-photon dissociation of CrO, directly analogous to the process discussed for MoO in Section IIIA. Figure 7 shows the total kinetic energy release for each channel observed as a function of photon wavenumber assuming an O atom co-fragment. A total of eight distinct spectral features are observed (labelled A-H in Figures 6 & 7), but not all features are evident in all images.

Assigning these series to specific dissociation channels proved more complicated than in the case of MoO, due to the abundance of energy levels arising from the Cr $3d^5\,4s^1$ configuration in the region explored. Literature values for $D_0$(CrO),[24-29, 39-41] are markedly lower than for MoO, leading to a wide range of accessible Cr* terms 20000 – 30000 cm$^{-1}$ above the ground state.[38, 42]

In this case, to aid assignment, each TKER spectrum in this spectral region was fit to a sum of Gaussian peaks corresponding to dissociation into the energetically allowed Cr* + O $^3P$ channels. With the TKER positions so fixed, the FWHM of each peak was determined by the spread of the spin-orbit components of each energy term and only the amplitude used as a fitting parameter. For each TKER spectrum the value for $D_0$ used to calculate the peak positions was varied until a reasonable fit was obtained.

The result of this process for one image (recorded at 21272 cm$^{-1}$) is shown in Figure 8, with the fit assuming a dissociation energy of $D_0$(CrO) = 37000 cm$^{-1}$. Similarly good fits are obtained for all other images. In many cases multiple Cr* terms lie sufficiently close enough together that individual channels cannot be assigned unambiguously. In such cases, a combination of all of the possible candidates was assumed, with the average of the appropriate energy levels used for the internal energy of the Cr fragment in calculating $D_0$, with the uncertainty increasing accordingly.

In this way it is possible to assign the eight series identified as follows:

- Series A is assigned to the Cr* $a\ ^3D\ (3d^54s)$ term at *ca.* 28670 cm$^{-1}$.

- Series B and C lie very close together in a region with several plausible assignments for Cr* = $b\ ^3P\ (3d^54s)$ at *ca.* 27180 cm$^{-1}$; $z\ ^7D\ (3d^44s4p)$ at *ca.* 27500 cm$^{-1}$; $b\ ^3G\ (3d^44s^2)$ at *ca.* 27700 cm$^{-1}$; and $y\ ^7P\ (3d^4\ 4s4p)$ at *ca.* 27800 cm$^{-1}$.

- Series D is assigned to Cr* = $a\ ^3G\ (3d^54s)$; $a\ ^3F\ (3d^44s^2)$; and $z\ ^7F\ (3d^44s4p)$ terms at 24900, 25100, and 25250 cm$^{-1}$, respectively.

- Series E is assigned to Cr* = $a\ ^3H_J\ (3d^44s^2)$ and $b\ ^5D_J\ (3d^54s)$ terms at 24050 and 24290 cm$^{-1}$, respectively.

- Series F is assigned to the Cr* = $z\ ^7P_J\ (3d^5\ 4p)$ and $a\ ^3P_J\ (3d^4\ 4s^2)$ terms at 23400 and 23600 cm$^{-1}$, respectively.

- Series G is assigned to the Cr* = $a\ ^5P\ (3d^5\ 4s)$ term at 21850 cm$^{-1}$.

- Series H is assigned to Cr* = $a\ ^5G\ (3d^5\ 4s)$ at 20520 cm$^{-1}$, the lowest energy chromium state in this region, the next lying 12500 cm$^{-1}$ lower, consistent with the rings in Series H being the largest TKER rings observed in this region.

Assuming these assignments, the five series D–H were used to find independent values for the dissociation energy using the same processes as in the preceding two sections, and these were averaged to produce $D_0$(CrO) of 37197 ± 78 cm$^{-1}$ (see Table 2 which also provides a summary of previous determinations). This is significantly lower than the value obtained for MoO which is consistent with the trends in the dissociation energy of other M–X bonds.[43] This value falls within the (wide) range of values from previous experimental studies but is in very good agreement with the guided ion beam study by Georgiardis and Armentrout.[26] It also agrees well with values from a range of computational studies, which fall in the region around 36400 – 41100 cm$^{-1}$.[7, 44-46]

### D. Dissociation of CrO in the Ultraviolet Region

Ultraviolet photodissociation of CrO region between *ca.* 230 and 245 nm resulted in very rich velocity map images, the assignment of which proved to be extremely challenging due to the very large number of possible terms for the Cr* fragment. Each individual velocity map image is a complex superposition of partially resolved rings and only when the data is plotted together, as in Figure 9(a), do several distinct, persistent series become clear.

The (common) slopes of the series present do not provide for an unambiguous assignment of the carrier with either CrO or Cr$_2$ plausible. However, the wide range of dissociation channels involved strongly suggests the former as single–photon dissociation of the closed shell, $^1\Sigma_g^+$ ground state of $Cr_2$[37] would be expected to produce a limited range of product quantum states. The more likely assignment is to two-photon dissociation of CrO (X $^5\Pi$ – Ref [47]). Figure 9(b) shows the TKER spectrum (assuming Cr* + O) at 42918 cm$^{-1}$, together with simulation of the available open dissociation channels, assuming this assignment and a CrO dissociation energy of 37000 cm$^{-1}$.

Given the difficulty of formulating a firm assignment for these data, the full process outlined in the previous section was not used to determine $D_0$(CrO). Instead, the eight most clearly resolved series observed were used to determine a best estimate of dissociation energy of CrO of 37200 cm$^{-1}$, consistent with the 37197 ± 78 cm$^{-1}$ reported in Section IIIC.

*IV. Conclusions*

Multiphoton dissociation of MoO and CrO has been investigated in both the visible and UV regions with velocity map imaging. Using isotope–selective fragment detection, the variation in the images with photon energies allows the dissociated species to be identified with confidence despite the lack of parent molecule selectivity. From the results in the visible region, dissociation energies of MoO and CrO of 44064 ± 133 cm$^{-1}$ and 37197 ± 78 cm$^{-1}$, respectively, have been determined. The additional experiments in the ultraviolet region support these values. In the case of MoO these results are consistent with the conclusions of recent high-resolution photoelectron studies / thermodynamic cycles. For CrO, where less of a consensus has emerged in the literature, this first determination using VMI supports the values derived from guided ion beam mass spectrometric studies.

**FIGURES:**

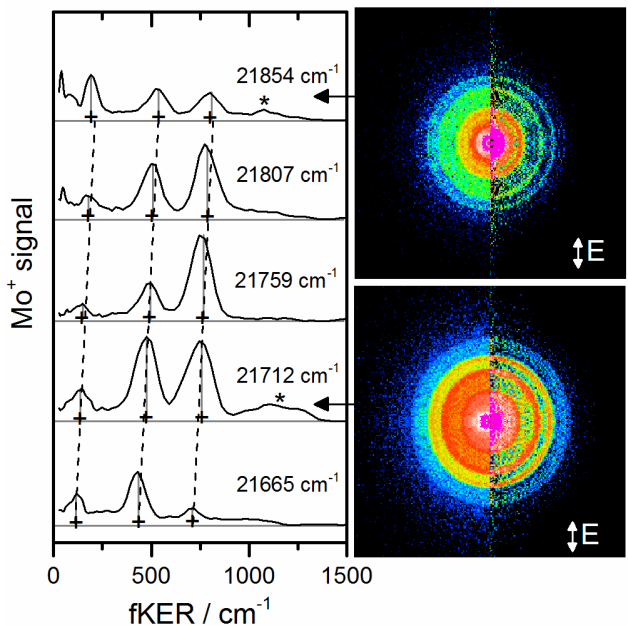

**Figure 1.** Mo* atom photofragment kinetic energy release spectra extracted from images recorded at different photon wavenumbers in the visible region. Representative velocity map images recorded at 21712 cm$^{-1}$ and 21854 cm$^{-1}$ are shown; the left half of each image shows the raw (symmetrised) data, while the right half shows the reconstructed central slice. Features marked with an asterisk are the weaker features shown in grey in Figure 2.

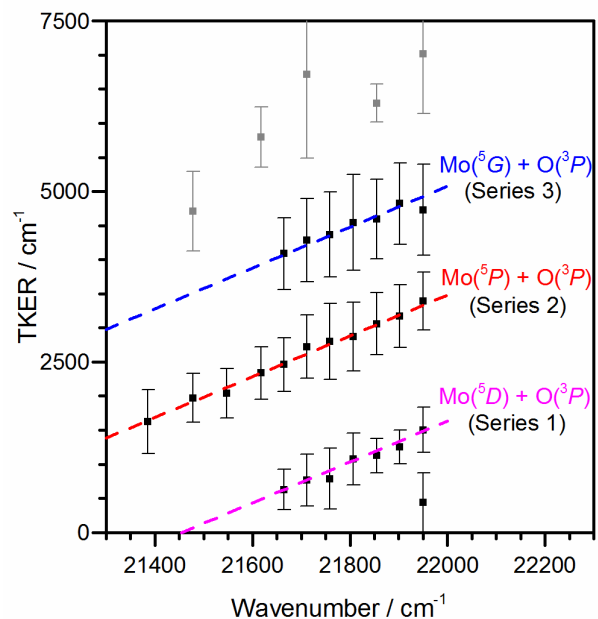

**Figure 2.** Variation with photon wavenumber of the main features in the kinetic energy release spectra observed following three-photon dissociation of MoO in the visible. Uncertainties indicate the full width at half maximum of the corresponding peaks in the TKER spectra. Points in grey are weaker features which have not been assigned.

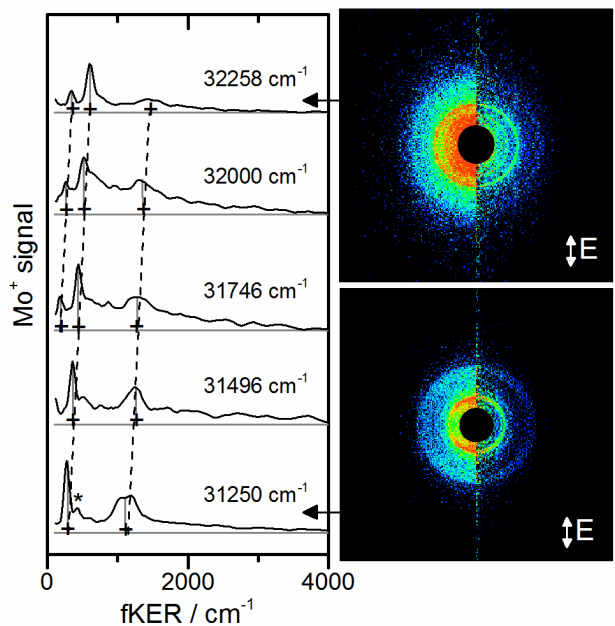

**Figure 3.** Mo* atom photofragment kinetic energy release spectra extracted from images recorded at different photon wavenumbers in the ultraviolet region. Representative velocity map images recorded at 31250 cm$^{-1}$ and 32258 cm$^{-1}$ are shown; the left half of each image shows raw (symmetrised) data, while the right half shows the reconstructed central slice. The feature marked with an asterisk is the feature shown in grey in Figure 4.

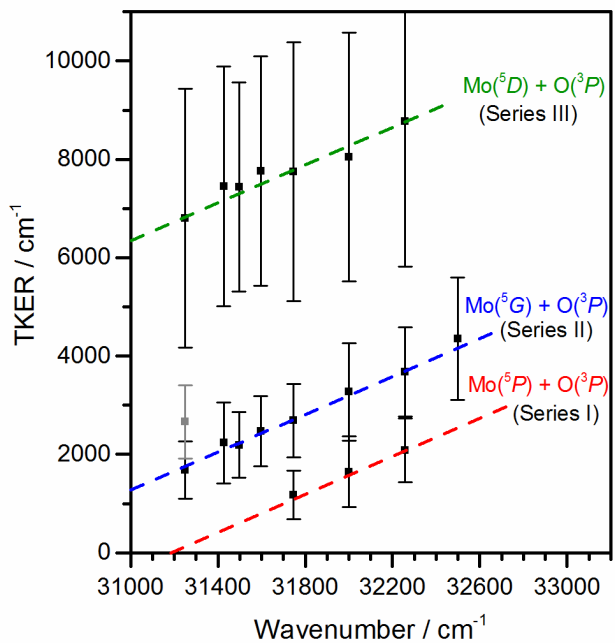

**Figure 4.** Variation with photon wavenumber of the main features in the kinetic energy release spectra following two-photon UV dissociation of MoO. Uncertainties indicate the full width at half maximum of the corresponding peaks in the TKER spectra. The point in grey indicates a weaker feature which has not been assigned.

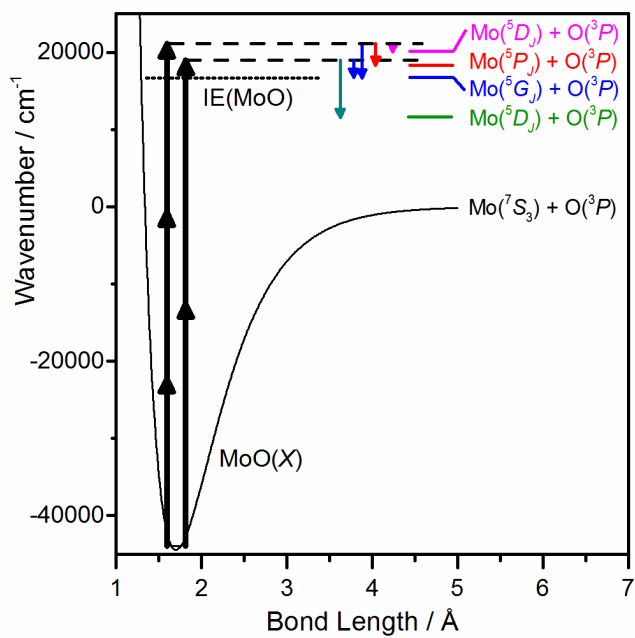

**Figure 5.** Schematic representation of the MoO ground state and relevant dissociation thresholds observed. Both two and three photon dissociation processes are shown. The dotted line indicates the MoO ionization energy.

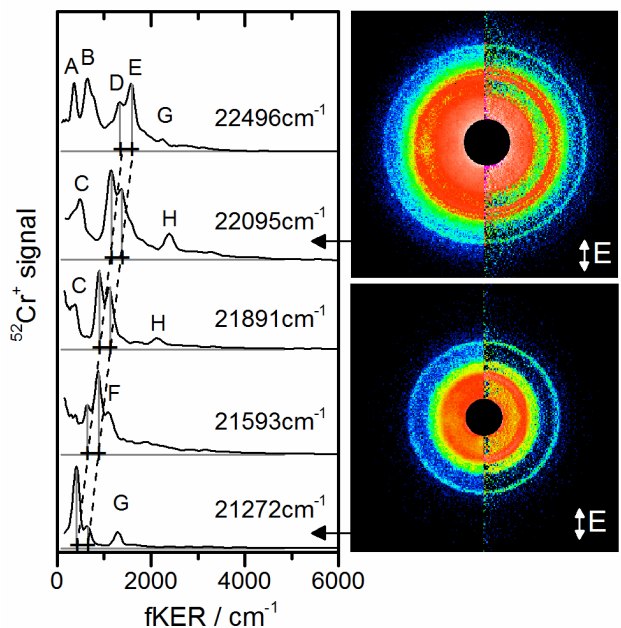

**Figure 6.** Cr* atom photofragment kinetic energy release spectra recorded at different photon wavenumbers in the visible region. Representative velocity map images are also shown; the left half of each image shows raw data, while the right half shows the reconstructed central slice. Letters A-H indicate different series observed (see Figure 7).

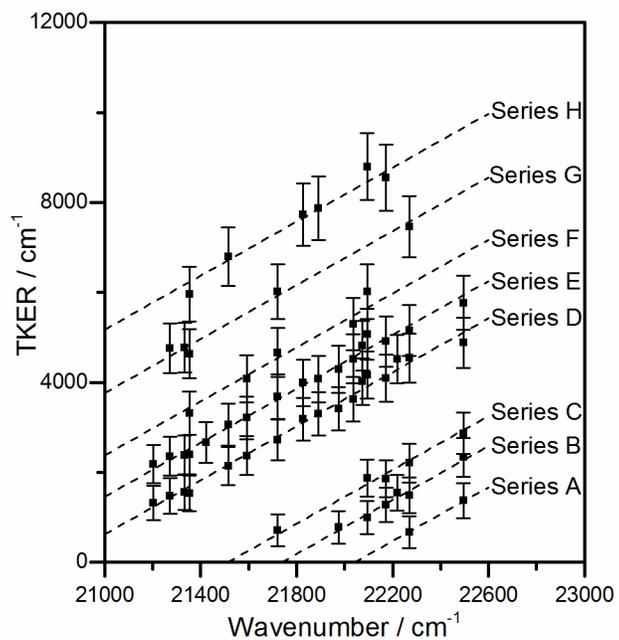

**Figure 7.** Variation with photon wavenumber of the principal features in the kinetic energy release spectra following three-photon dissociation of CrO. Uncertainties indicate the full width at half maximum of the corresponding peaks in the TKER spectra.

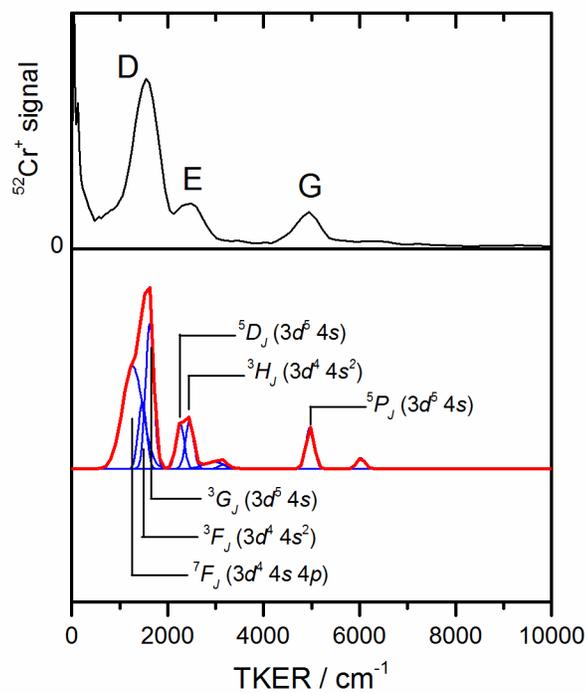

**Figure 8.** (upper) Experimental kinetic energy release spectrum recorded at 21272 cm$^{-1}$ following three-photon dissociation of CrO. (lower) Fitted TKER spectrum in which the blue lines indicate Gaussian peaks corresponding to individual Cr* + O $^3P$ dissociation channels with peak widths calculated from the spread of the spin-orbit levels. The red line is a cumulative fit. A dissociation energy of $D_0$ = 37000 cm$^{-1}$ is assumed.

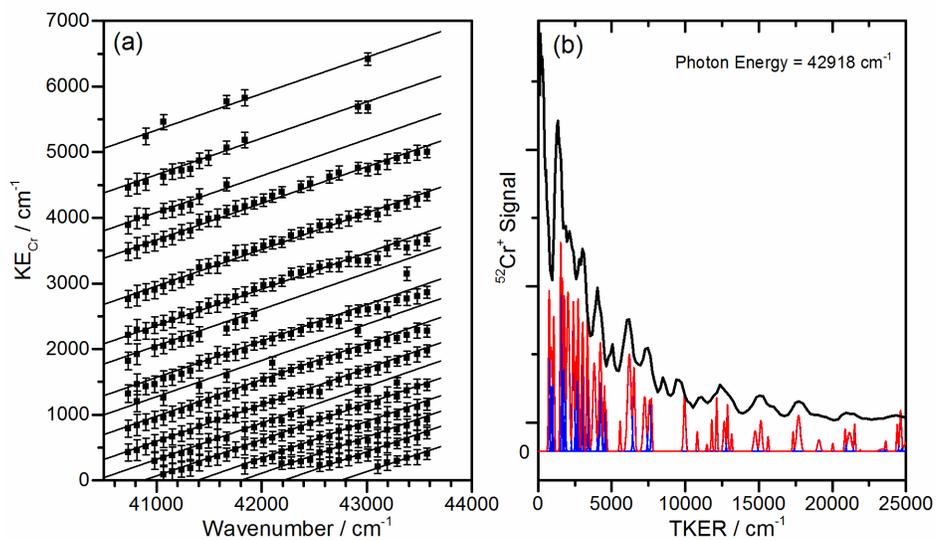

**Figure 9.** (a) Variation with photon wavenumber of the main features in the kinetic energy release spectra following two-photon (UV) dissociation of CrO. Uncertainties indicate the full width at half maximum of the corresponding peaks in the TKER spectra. (b) An experimental TKER spectrum recorded at 42918 cm$^{-1}$. The black line is the experimental data; blue lines indicate Gaussian peaks corresponding to individual Cr* + O $^3P$ dissociation channels and the red line is a cumulative fit assuming a dissociation energy of 37000 cm$^{-1}$.

**Table 1**: Comparison of the MoO ground state dissociation energy determined in this work with that from other previous studies.

| $D_0$ /cm$^{-1}$ measurement 1: (Visible) | $D_0$ /cm$^{-1}$ measurement 2: (UV) | Expt./eV | $D_e$ Calc./eV |
|---|---|---|---|
| 44064 ± 133[a] | 43895 ± 148[b] | 5.44 ± 0.04 (43900 ± 300 cm$^{-1}$)[g] | 5.58 (45000 cm$^{-1}$)[i] |
| 44069 ± 265[d] | 43945 ± 1099[c] | 5.42 ± 0.02 (43700 ± 200 cm$^{-1}$)[h] | |
| 44027 ± 217[e] | 43905 ± 190[d] | | |
| 44100 ± 222[f] | 43875 ± 243[e] | | |

[a] Weighted-average of all visible values
[b] Weighted-average of all ultraviolet values
[c] Derived from the Mo* $^5D$ ($4d^4 5s^2$) + O $^3P$ channel data in isolation
[d] Derived from the Mo* $^5G$ + O $^3P$ channel data in isolation
[e] Derived from the Mo* $^5P$ + O $^3P$ channel data in isolation
[f] Derived from the Mo* $^5D$ ($4d^5 5s$) + O $^3P$ channel data in isolation
[g] Loock et al.[8]
[h] Luo et al.[9]
[i] $D_e$ from Oliveira et al.[48]

**Table 2**: Comparison of values for the ground state experimental dissociation energy, $D_0$ /cm$^{-1}$, for CrO (from this work and previous studies).

| Year | Author | $D_0$ / cm$^{-1}$ | Experimental Method |
|---|---|---|---|
| 1932 | Ghosh[39] | 30500 | Electronic Spectroscopy |
| 1952 | Huldt & Lagerqvist[40] | 35550 ± 3900 | Flame Spectroscopy |
| 1961 | Grimley *et al.*[41] | 35410 ± 2000 | Knudsen Cell |
| 1981 | Balducci *et al.*[24] | 36510 ± 750 | Knudsen Cell |
| 1986 | Kang & Beauchamp[25] | 39000 ± 700 | Guided Ion Beam |
| 1989 | Georgiadis & Armentrout[26] | 36900 ± 600 | Guided Ion Beam |
| 1993 | Ebbinghaus[27] | 38800 | Thermodynamic Cycle |
| 1996 | Hedgecock *et al.*[28] | 38200 ± 700 | Crossed Molecular Beam |
| 2010 | Smirnov & Akhmadov[29] | 38510 | Reaction Kinetics |
| 2017 | This work | 37197 ± 78 | Velocity Map Imaging |

**AUTHOR INFORMATION**

**Corresponding Author**

*stuart.mackenzie@chem.ox.ac.uk

**ACKNOWLEDGMENTS**

G.A.C. is grateful to EPRSC and Magdalen College Oxford, for his graduate studentships. A.I. is similarly grateful to EPSRC and Wadham College, Oxford. A. S. G. is grateful to the EPSRC for funding his postdoctoral position. This work is supported by EPSRC under Programme Grant No. EP/L005913.